# Hybrid Nanofluids: An Overview of their Synthesis and Thermophysical properties.

**Ansab Jan[1], Burhan Mir[1], Ahmad A Mir[1]**

[1]Mechanical Department, National Institute of Technology Srinagar, J & K State, India

Email:   janansab@gmail.com, burhanmir137@gmail.com, ahmadmir96@gmail.com

Phone: (+91)7006022162, (+91)9906428834, (+91)9858016267

**Abstract.** A recent area of interest in thermal applications has been nanofluids, with extensive research conducted in the field, and numerous publications expounding on new results coming to the fore. The properties of such fluids depend on their components, and by the use of combined and complex nanoparticles with diverse characteristics, the desirable qualities of multiple nanoparticle suspensions can be obtained in a single mixture. The paper covers these hybrid nanofluids, the techniques for their synthesis, their properties, characteristics, and the areas of debate and outstanding questions.

## 1. Introduction

Nanofluids consist of nanosized particles suspended in a liquid, and have shown remarkable potential for enhancement in thermal properties with respect to their base fluids. These are the next generation of working fluids, set to replace of their conventional counterparts. A recent update to the nanofluid scene, hybrid nanofluids, with suspended particles being complex combination of multiple nanoparticles, aim to rectify the shortcomings of mono nanofluids by incorporating a contrasting property additive to offset any disadvantages of the mono nanofluid. Along with commonly used particles, metals and their crystalline oxides, ceramic oxides, and carbon based particles (CNTs, graphene) can be used in combinations, for achieving desired output characteristics.

## 2. Synthesis

The synthesis of nanoparticles and the subsequent preparation of nanofluids is carried out by different methods depending on the type of nanoparticles used. Han *et al.* [1] used a process to create hybrid sphere/carbon nano tube (CNT) nanoparticles, to reduce thermal contact resistance between CNTs by providing the thermal contact via central spheres of crystalline metal oxides. The spherical nanoparticles were produced in two steps: spray pyrolysis of the metal salt solutions, followed by catalytic growth of CNTs at these particle surfaces. An aqueous precursor solution of $Fe(NO_3)_2$ and $Al(NO_3)_2$ with mixing ratio 1:1 of concentration 3 wt.% was produced. Aerosol droplets of the solution were decomposed in nitrogen carrier gas to form Fe and Al nitrate nanoparticles. Water was removed in a silica gel dryer, and the mixture, with hydrogen, was introduced to a tube furnace for pyrolytic decomposition to crystalline oxide nanoparticles. In another tube furnace, the particles react with acetylene and hydrogen at 750°C resulting in the growth of CNT at their surfaces. The end resultant has mass fractions of individual components as CNT =3 wt%, alumina=32 wt%, iron oxide=65 wt%. These particles are then dispersed to polyalphaolefin (PAO) oil with sonication and a suitable surfactant, forming the hybrid nanofluid.

Jana *et al.* [2] synthesized hybrid nanofluid containing CNT, gold nanoparticle (AuNP) and copper nanoparticle (CuNP). CNTs were polarized by chemical treatment for better dispersal. 1 gram of CNT was suspended in 40ml mixture of concentrated nitric acid and sulphuric acid (1:3 v/v) and refluxed at 140°C for 1 hour. CNTs were filtered and washed with deionized water until pH reached around 7. Soaked CNTs were dried in vacuum oven at 150°C for 12 hours. CNTs in different volume fraction (0.2, 0.3, 0.5, 0.8%) were added to water to produce suspensions. AuNP colloid was then added to deionized water in the ratio of 1.4:1. AuNP suspensions were added to CNT suspension having different volume fraction of CNTs, in 1.5:2.5 ratios to achieve CNT-AuNP suspension. The CuNP suspensions composed pf CuNPs, laurate salt (9% by wt.) and deionized water (volume fraction 0.05, 0.1, 0.2, 0.3%). Each type of CuNP was added to CNT (0.5 vol%) in 1.5-2.5 ratios.

Ho *et al.* [3] synthesized hybrid nanofluid containing micro encapsulated phase change material (MEPCM) or n-eicosane. PCM suspension is prepared by interfacial polycondensation with emulsion technique, n-eicosane was emulsified in water-soluble urea-formaldehyde per-polymer solution without deliberately added emulsifier. PCM is about 60% by wt. of MEPCM particle. PCM solution was formulated by mixing appropriate quantities of MEPCM with ultra-pure Milli-Q water in a flask, dispersed in ultrasonic vibration bath for 2 hours (Mass fraction's:3.7, 9.1, 18.2 wt%. Further, $Al_2O_3$ nanoparticles of various mass fractions (2-10 wt%) were dispersed in ultra-pure water using magnetic stirrer for more than 4 hours and the pH was adjusted to about 3. Hybrid nanofluid was obtained by mixing the nanofluid with PCM suspension in an ultrasonic vibration bath.

Suresh *et al.* [4] used ceramic materials like Alumina ($Al_2O_3$) having excellent stability, but lower thermal conductivity when compared to their less stable metallic counterparts. The idea was to incorporate small amounts of Cu into the alumina matrix, to improve the thermal properties, with minimal effect on the nanofluid stability. A solution of the salts Cu $(NO_3)_2.3H_2O$ and $Al(NO_3)_3 \cdot 9H_2O$ was first prepared in water and then spray dried at 180°C. The powder obtained was heated at 900°C for about 60 min, to form powdered copper oxide and alumina. The mixture was heated in hydrogen rich environment at 400°C for an hour, preferentially reducing CuO to Cu. The resultant was ball milled and then dispersed in deionized water with sodium laurel sulfate (dispersant) using an ultrasonic vibrator.

Paul *et al.* [5] synthesized Al-Zn based hybrid nanofluid. Nanoparticles were prepared by mechanical alloying, allowing production of homogeneous powdered mass of ultra-fine or nano-sized particles with polycrystalline micro-structure. The constituent powders were mixed in proportion, loaded along with grinding balls in closed containers and placed on the sun disc of a planetary mill. Elemental powders of Al and Zn were taken and a blend of Al and 5 wt% Zn was subjected to mechanical alloying at room temperature using high energy planetary ball mill at 300 rpm and 10:1 ball to powder weight ratio. The powder blend was milled until desired composition was achieved. The nanofluids were prepared by two-step process by adding ultra-fine Al and 5 wt% Zn nanoparticles in appropriate quantity of ethylene glycol and subjecting the resultant mixture to ultrasonic vibration and magnetic stirring for to ensure stable dispersion and isothermal condition in base fluid.

Afrand *et al.* [6] took nanoadditives, composed of equal volumes of functionalized multi walled carbon nanotubes (FMWCNTs) and MgO nanoparticles. These were dispersed in an ethylene glycol base. The selected particles have contrasting favourable thermal properties.

Asadi *et al.* [7] prepared an $Al_2O_3$-MWCNT (85%-15%) hybrid nanofluid in oil with solid concentrations-0.125, 0.25, 0.5, 0.75, 1 and 1.5%. MWCNTs used had an outer diameter of 20-30 nm, inner diameter of 5-10 nm and a length of 10-30 nm. The size of $Al_2O_3$ particles was 20 mm. A magnetic stirrer was employed for 2 hours after which an ultrasonic homogenizer was used for 1 hour in order to break down the agglomeration of particles.

Botha *et al.* [10] used a one-step method to prepare a nanofluid with Ag nanoparticles supported on silica. Silica taken in concentrations between 0.07wt% and 4.4wt% was mixed in transformer oil followed by magnetic stirring for 2h at 130°C. Thermal conductivity measurements were carried out before settling occurred due to the poor stability of the silica nanofluids in the absence of surfactant. Silver nanoparticles (0.1wt% to 0.6wt %) supported on silica were prepared in a similar way by introducing the silver nitrate and silica to the base fluid. The temperature was raised to 130°C for 2h to cause the oil to oxidise and hence lead to the reduction of $Ag^+$ ions to Ag by electron transfer reaction. Finally, the mixture was brought down to room temperature and the nanofluid was structurally characterized using UV-via TEM, and XRD.

In another experiment, Afrand *et al.* [11] prepared $Fe_3O_4$-Ag/EG hybrid nanofluid samples, with the solid volume fractions of 0.0375%, 0.075%, 0.15%, 0.6% and 1.2%, by a two-step method. In this process equal volumes of Fe3O4 and Ag nanoparticles were dispersed in EG. In order to prepare stable samples, after magnetic stirring for 2.5h, the suspensions were exposed to an ultrasonic processor (Hielscher Company, Germany) with the power of 400W and frequency of 24 kHz for 6h. This helped in breaking down the agglomerated particles.

In the experiment conducted by Esfe *et al.* [13], the properties of Ag-MgO/Water hybrid nanofluid were studied. The nanoparticles were dispersed in a base fluid using a two step method. The mixture was then placed in an ultrasonic vibrator for 3 hours. As a surfactant, low concentrations of Cetyl Trimethy Ammonium Bromide (CTAB) to ensure stability and proper dispersion. The concentration of the surfactant was kept low enough to not affect the nanofluids' thermophysical properties. The process was repeated for preparing different concentrations of the nanofluid.

## 3. Properties

In the case of hybrid sphere/CNT nanoparticles dispersed in PAO oil [1], the experimental evidence suggested an appreciable enhancement of thermal conductivity relative to the base fluid, and related mono nanofluids (figure. 1), primarily credited to diffusive heat conduction mechanism. An enhancement of 21% was observed with respect to PAO, at 0.2 vol. % and around half the value at 0.1 vol. %.

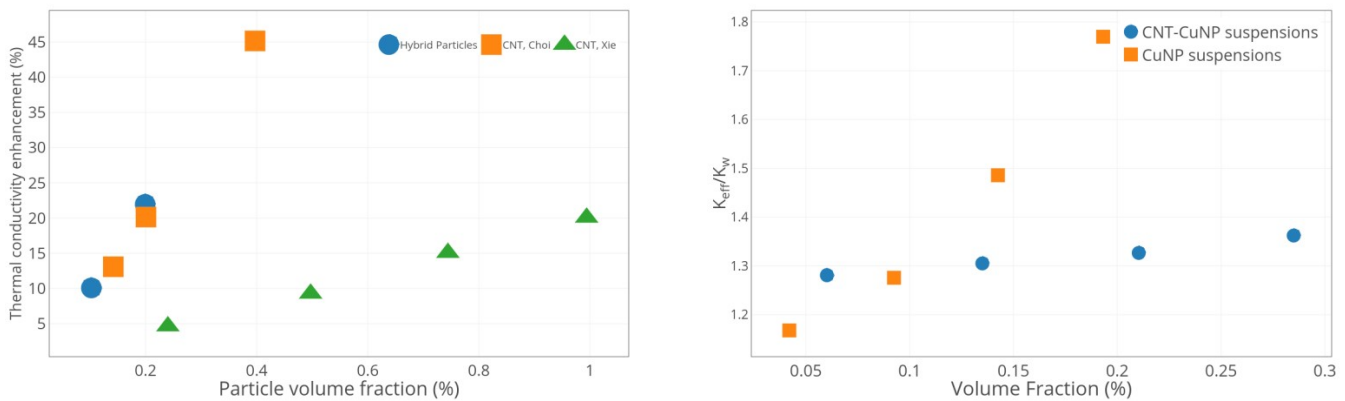

**Figure 1:** Performance comparison of particles with different morphologies:CNTs and hybrid sphere/CNT particles (urchin-like), in nanofluids. CNTs in PAO by Choi et al. [8] and

CNTs in decene by Xie et al. [9]
**Figure 2:** Comparison of normalized thermal conductivity between CuNP suspensions and CNTCuNP suspensions as a function of CuNP concentration [2]

Hybrid nanofluids containing CNT, AuNP and CuNP showed enhanced thermal conductivity compared to the base fluid as well as nanofluids containing only CNT or Cu NPs, shown in figures 2 and 3.

Ho *et al.* [3] discovered an increse thermal conductivity with increase in nanoparticle concentration at different concentrations of PCM (figure 4). Dynamic viscosity of the nanofluid without PCM increased slightly with increase in mass fraction of nanoparticles as compared to pure water.

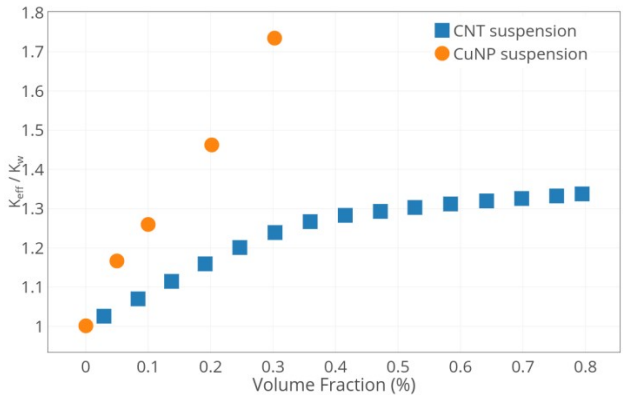

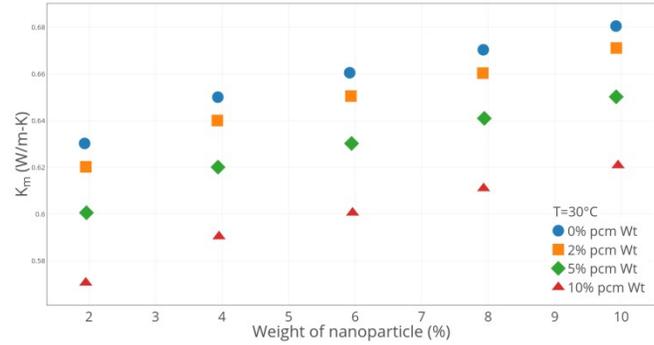

**Figure 3:** Normalized thermal conductivity of CNT and CuNP suspensions as function of concentration [2]

**Figure 4:** Thermal Conductivity measured for the hybrid water based suspensions of Al2O3 nanoparticles and MEPCM particles [3]

Suresh *et al.* [4] concluded that the thermal conductivity of the hybrid nanofluid increases almost linearly with increase in the volume concentration of the nanoparticles. The data has been represented in figure 5, where a comparison with alumina/water nanofluid is also shown. Hybridization of the nanofluid resulted in significant enhancement of thermal conductivity. For all volume concentrations in the study, the fluid remained Newtonian. The viscosity increased with particle volume concentration. The difference in viscosity between $Al_2O_3$-Cu/water and $Al_2O_3$/water increased with nanoparticle concentration.

Paul et al. [5] employed transient hot wire method to measure thermal conductivity. The thermal conductivity increased linearly for both the base and nanofluids, with the increase being greater in higher concentration nanofluids, as seen in figure 6.

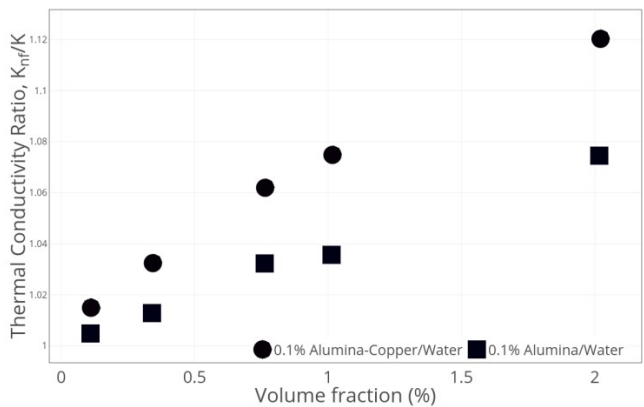

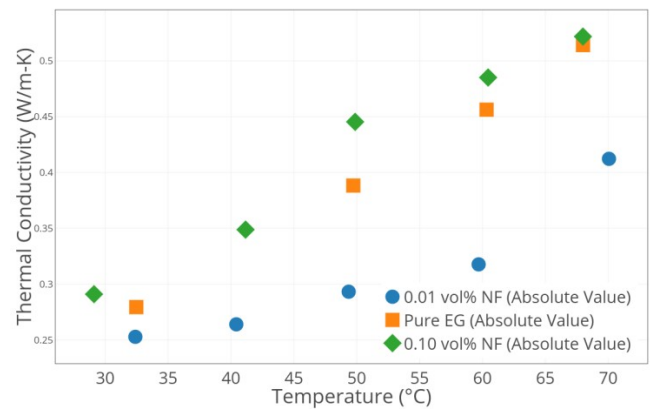

**Figure 5:** Thermal conductivity of Al$_2$O$_3$-Cu/water [4]

**Figure 6:** Thermal conductivity of Al-5wt%Zn dispersed ethylene glycol based nanofluid as a function of temperature [5]

Similar to previous observations, in case of the MgO-FMWCNT/ EG nanofluid, Afrand [6] found that the thermal conductivity increased with increase in volume concentration of solid, mainly due to Brownian movement and an increase in the level of interactions between the nanoparticles. Thermal conductivity also increased with temperature, with the rate of increase being greater at lower concentrations, probably due to larger clusters.

A comparison with certain monofluids as shown in figure 7 indicates that the thermal conductivity of the hybrid nanofluid was enhanced to a greater degree, due to the greater inherent thermal conductivity (in most cases). When compared to CNT/EG nanofluid, addition of MgO in the hybrid nanofluid prevents clustering in the fluid.

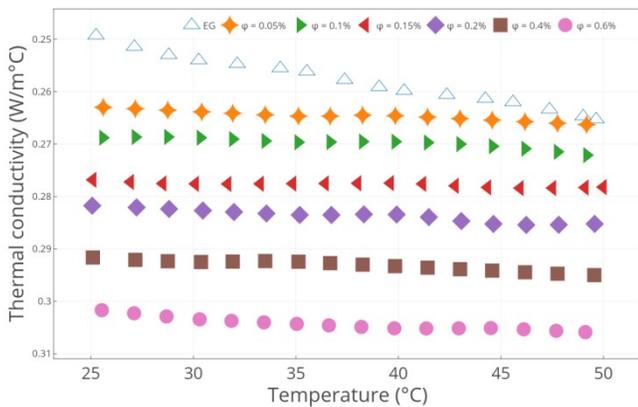

**Figure 7:** Variations of thermal conductivity vs temperature for various nanofluid samples [6]

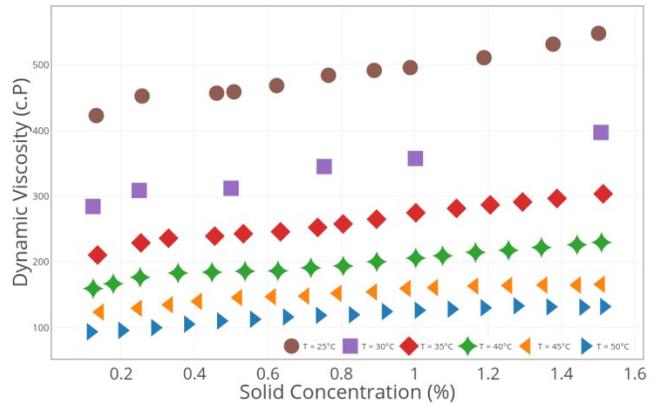

**Figure 8:** Dynamic viscosity with respect to solid concentrations [7]

Asadi *et al.* [7] found that the hybrid nanofluid of Al$_2$O$_3$ MWCNT in oil showed Newtonian behaviour under the conditions maintained for the study. Thermal conductivity enhancement occured with increase in temperature and particle concentration. Dynamic viscosity increased as solid concentration increased, shown in figure 8. Heat transfer efficiency of the hybrid nanofluid with respect to base oil was better for internal laminar flow at all concentrations. The same was better for internal turbulent flow at concentrations lower than 1%.

Figure 9 by Botha *et al.* [10] shows the thermal conductivity of silica with concentrations ranging from 0.07 wt % to 4.4 wt %. An increase of 1.7% in thermal conductivity was observed with 0.5 wt% silica and that of 3.5% for 1.8 wt% silica, without silver nanoparticles, in oil. The highest concentration of silica used was 4.4 wt% it showed an increase of 5.2% in thermal conductivity (figure 10). However, the higher the concentration of silica, the more gel-like the suspensions, which indicates that the main mechanism for heat transfer is via particle-particle contact, and not Brownian motion. Thus, there is small increase in thermal conductivity at low wt% as compared to higher wt%. When Ag was supported on silica, the thermal conductivity was found to increase with an increase in silver concentration (Figure 11). A significant improvement in thermal conductivity (15%) was observed when only 0.60 wt % Ag was supported on 0.07 wt % silica. The silver nanoparticles show enhanced thermal properties possibly due to the particles being close enough for thermal transport to take place, and suitably supporting the particles provides for a stable heat transfer system.

Using the Brookfield DV-I PRIME digital Viscometer equipped with a temperature bath, viscosity measurements were performed by Afrand *et al.* [11] at temperatures of 25, 30, 35, 40, 45 and 50°C in the shear rate range of $12.23 s^{-1}$ to $122.3 s^{-1}$. All the experiments were repeated at different shear rates for all samples at each temperature in order to determine the Newtonian or non-Newtonian behavior of the nanofluid samples. A comparison was made between the viscosities of EG obtained by the Viscometer at different temperatures and those reported by Chen *et al.* [12]. Little difference was found between the two (maximum 2.67%).

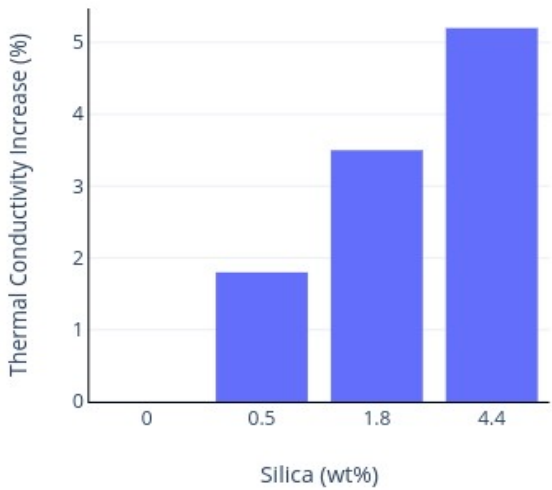

**Figure 9:** Thermal conductivity of silica with concentrations ranging from 0.07 wt % to 4.4 wt % [10]

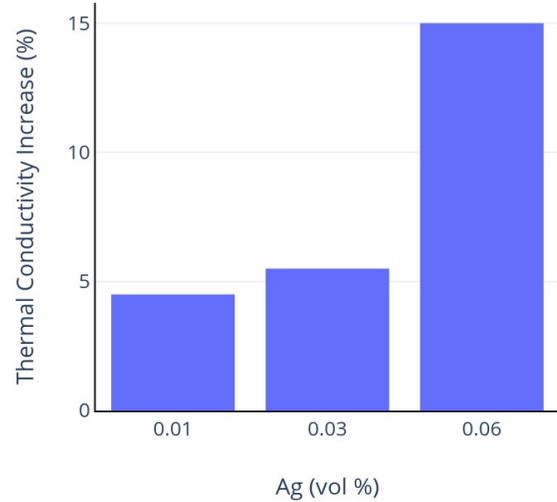

**Figure 10:** Thermal conductivity increase as a function of Ag concentration, supported on 0.07 wt % silica. Because of the poor dissolution of silver salt in oil, a maximum concentration of 0.60 wt % silver was used [10]

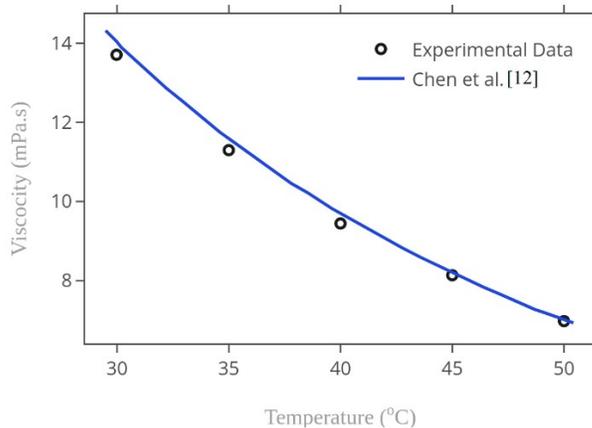

**Figure 11:** Comparison between measurements and data obtained by Chen et al. [12] for EG at different temperatures [11]

The following equation was obtained through curve fitting of experimental data (Figure 12) for thermal conductivity of nanofluds at different volume fractions by Esfe *et al.* [13]:

$$k_{nf} = \left( \frac{0.1747 \times 10^5 + \Phi_p}{0.1747 \times 10^5 - 0.1498 \times 10^6 \Phi_p + 0.1117 \times 10^7 \Phi_p^2 + 0.1997 \times 10^8 \Phi_p^3} \right) k_f$$

$0 \leq \Phi \leq 0.03$

The thermal conductivity ratio of the hybrid nanofluid and base fluid was calculated using the proposed model for concentrations between 0 and 2%, and the information was compared with various previous models. The models used for the comparison were Hamilton and Crosser model [14], Yu and Choi model [15], and the first and second models of Prasher et al. [16]. The comparison is shown in figure 13.

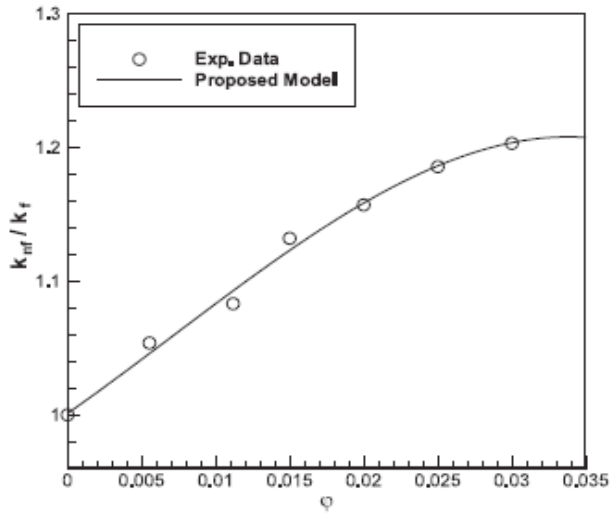

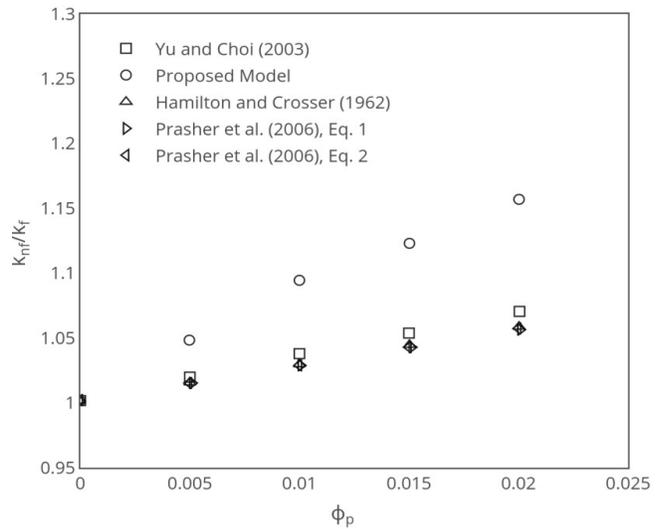

**Figure 12:** Curve fitting on experimental data of nanofluid thermal conductivity [13]

**Figure 13:** Comparison between the prediction thermal conductivity from different models (correlations) with proposed correlation [14]

From the figure we can draw the conclusion that these models underestimate the thermal conductivity increase when compared to the correlation obtained, and with increased nanoparticle volume concentration, the difference becomes more significant. Table 1 shows a numerical comparison shows a comparison between measured valued of thermal conductivity ratios by Esfe et al. [13] with the values predicted by various models. The increase in gradient can clearly be seen in the table with the increase in nanoparticle concentration.

| Nanoparticle volume fraction | 0 | 0.005 | 0.01 | 0.015 | 0.02 |
|---|---|---|---|---|---|
| Presented Model | 1 | 1.047 | 1.094 | 1.124 | 1.158 |
| Yu and Choi [?] (%diff) | 1 | 1.017 (2.8%) | 1.034 (5.4%) | 1.053 (6.3%) | 1.071 (7.6%) |
| Hamilton and Crosser [?] (%diff) | 1 | 1.014 (3.1%) | 1.029 (6%) | 1.044 (7.1%) | 1.059 (8.6%) |
| Prasher et al. [?], Eq. (1) (%diff) | 1 | 1.014 (3.1%) | 1.029 (6%) | 1.044 (7.1%) | 1.059 (8.6%) |
| Prasher et al. [?], Eq (2) (%diff) | 1 | 1.014 (3.1%) | 1.029 (6%) | 1.044 (7.1%) | 1.059 (8.6%) |

Table 1: Numerical comparison of measured values of thermal conductivity ratio with predicted value by different thermal conductivity ratios models.

| Nomenclature | |
|---|---|
| PAO | Polyalphaolefin |
| CNT | Carbon Nanotube |
| MWCNT | Multi-walled carbon nanotube |
| FMWCNT | Functionalized multi-walled carbon nanotube |
| NP | Nanoparticles |
| MEPCM | Micro encapsulated change material |

## 4. Conclusion

From the analysis of the available data, it can be concluded that hybrid nanofluids have great potential in thermodynamic applications, with larger control over the behaviour of the manufactured fluid. With the option to meticulously engineer nanofluids with specific applications in mind, the advent of hybrid nanofluids has potential to drastically change the field of heat transfer. The challenges of economics and stability can be more effectively tackled which opens many new avenues for such products.